# *UWB patch antenna design and realization in the bandwidth 780MHz to 4.22GHz*

*Chemseddine BENKALFATE[(1)] and Mohammed FEHAM[(1)]*

*and Achour OUSLIMANI[(2)] and Abdelhak KASBARI[(2)]*

(1)STIC Laboratory, University of Tlemcen, Algeria,
(2)Quartz Laboratory, ENSEA, France
chemseddine.benkalfate@univ-tlemcen.dz

**Abstract.** The proposed UWB antenna covers mobile communications (GSM, EDG, UMTS(3G), LTE(4G) …) and wireless networks (WIFI, WiMAX), within a theoretical bandwidth defined from 780MHz to 4.22GHz. The UWB antenna is designed and realized on a FR-4 substrate with an electrical permittivity $\varepsilon_r$ of 4.4. It presents a 98.75% average analytical efficiency and an omnidirectional radiation within the previous bandwidth. The impedance excitation port is fixed at 50 Ohm according with the SMA impedance used in the practical part. The measured results are in good agreement with those obtained using CST and ADS software's. The measured bandwidth, defined from 980MHz to 4.2GHz, presents an efficiency of 94.14%. Furthermore, the practical radiation diagram and the excitation port impedance stay the same as that the simulation one.

## 1  Introduction

The communication systems communicate with others via a device called antenna. There are three major antenna types, the monomodal antenna [1], the multiband antenna [2] and the ultra-wideband antenna [3][4]. The first antenna type, works with a just one precise frequency, the second type can catch a finite number of waves at a precise frequency and the third antenna type offers a possibility to catch all waves ambient in a defined wideband. This paper is focused on the UWB antenna, which is derived from [5]. In this reference [5], the antenna is defined by a coplanar waveguide excitation, and is addressed to [4 – 10] GHz applications, so this antenna can't cover the mobile communications and wireless networks. Our purpose is to make this antenna able to be operational within the last networks bandwidth. For that, we brought a radical edit, which consists to change the coplanar waveguide to a microstrip line feeding, where the ground plan has a particular form getting after modification of the coplanar waveguide and it's printed in the bottom side of the substrate. In the other hand, the patch printed on the top side, has been also modified by increasing very carefully all parameters values and make the feeding line longer than what it was and this length has to be equal to $\frac{\lambda_0}{2}$, $\lambda_0$ is a wavelength of the central frequency band-width. With all these modifications, we've got an antenna frequency response going from [0.77 – 4.2] GHz, which is required. In the following, we present the proposed antenna structure and the CST simulation results and we deduce its electrical equivalent circuit which going to be simulated by ADS software. The experimental and simulation results are then compared. Finally, we give the realized antenna model and the practical performances [6].

## 2    The proposed UWB antenna simulation

In this part, the most important antenna parameters are simulated and analyzed to precise the proposed antenna performances.

### 2.1   The proposed antenna model

The proposed antenna model, is a patch antenna, such as the strip face contain six resonators, each one, radiates in a precise frequency. The thickness of this proposed antenna is 1.7 mm, the used substrate is the FR4($\varepsilon_r$=4.4), and the total width and height are 78 mm and 120.75 mm respectively. the ground plan has a particular form, presented in the Figure 1, and the strip side [6] is depicted in the Figure 2. This particular ground form permits to widen the frequency bandwidth, and pass from a multiband antenna to an UWB antenna generally. In this context, the dimension values of our proposed antenna, which give a perfect UWB response, are listed in the Table.1.

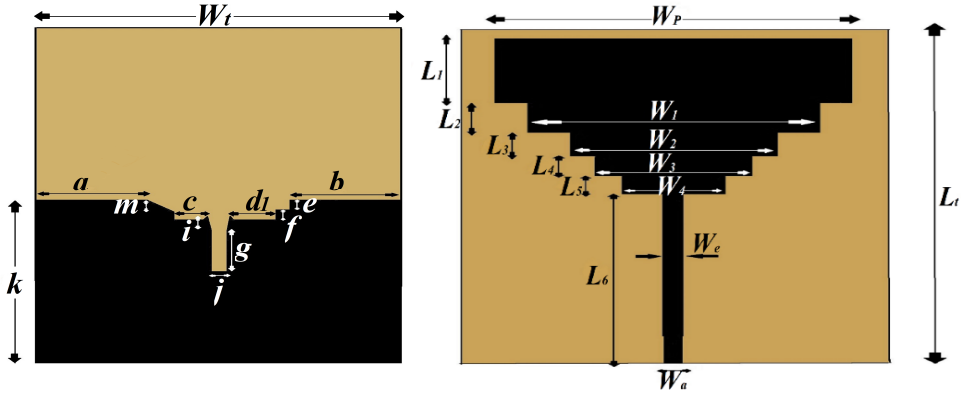

**Fig. 1.**   Antenna ground side   **Fig 2.**   Antenna top side

**Table. 1**.   Antenna dimension values

| *Dimensions* | $L_1$ | $L_2$ | $L_3$ | $L_4$ | $L_5$ | $L_6$ | $L_t$ | $W_1$ | $W_2$ |
|---|---|---|---|---|---|---|---|---|---|
| *Values(mm)* | 24 | 10.5 | 8.5 | 7.5 | 6.75 | 63 | *120.75* | 51 | 36 |
| *Dimensions* | $W_3$ | $W_4$ | $W_p$ | $W_e$ | $W_t$ | a | b | c | $d_1$ |
| *Values(mm)* | 27 | 18 | 62 | 3.7 | 78 | 24 | 24 | 7.1 | 9.66 |
| *Dimensions* | e | f | g | i | j | k | $W_a$ | m | |
| *Values(mm)* | 3.7 | 3.8 | 16 | 5.3 | 3.25 | 67.5 | *3.2* | 3.7 | |

## 2.2 Simulated return loss

The return loss parameters $S_{11}$ is too important for being able to extract the performance of this antenna. Its magnitude is given according with the incident and the reflected power in level to the first port of the antenna. The Figure 3 shows the representation of this antenna in term of the scattering parameters.

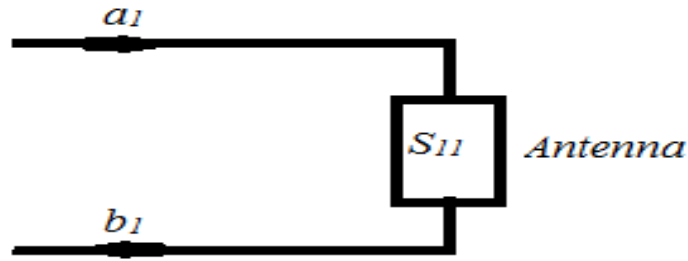

**Fig. 3.** Representation of the Scattering matrix

$a_1$ and $b_1$ are defined as the incident and the reflected power respectively. Then, the return loss is given by the following expression (equation 1).

$$s_{11} = \frac{b_1}{a_1}| \qquad (1)$$

Most of times we fix the minimum acceptable front $S_{11}$ level to -10 dB, however, when the reflected power rises up to the 10% of the incident power, this level of reflection is considered like a maximum acceptable reflection level, so a good antenna response is defined when the reflected power being less than the 10% of the incident power, in the other way, when $S_{11}$ become less than or equal to -10 dB
Our proposed antenna is simulated by CST software, the corresponding $S_{11}$ is depicted in the Figure 4.

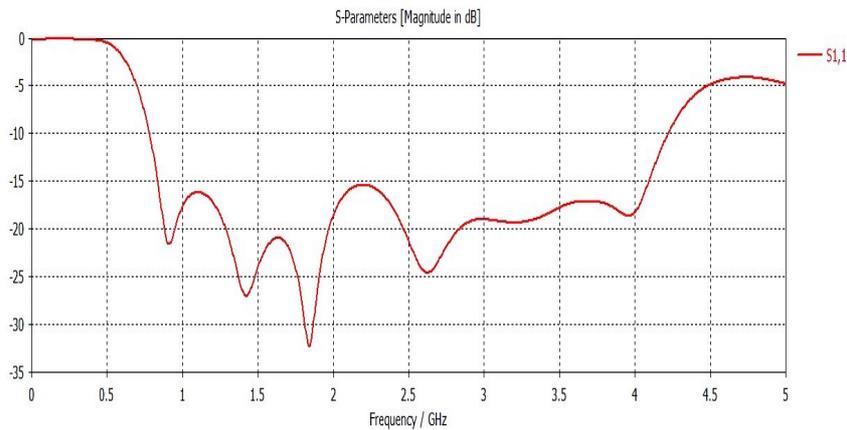

**Fig. 4.** Reflection coefficient variation

As we can see, the acceptable band is given according with the $S_{11}$ parameter, such as this favorable band starts from around 780 MHz to 4.2 GHz with an $S_{11}$ less than -15 dB, that means there are the minimum possible reflections in this bandwidth.

### 2.3 Voltage Standing Wave Ratio (VSWR)

In general case the VSWR has to be less than 2, that means almost 90% of the power has to be radiated and only nearly 10% of the power has to be reflected back. So, the VSWR value can be seen to be within 1 to 2 in the operating range. The simulation results for VSWR for the frequency range from [0.780 - 4.2] GHz is shown in the Figure 5.

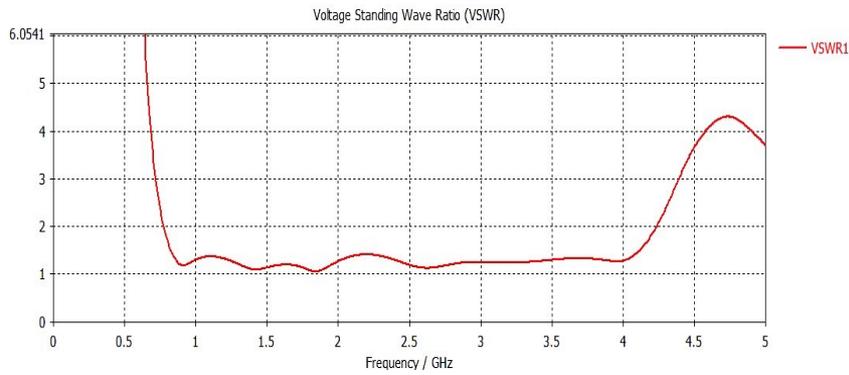

**Fig. 5.** VWSR antenna simulation

### 2.4 Radiation diagram

The radiation diagram shows in tridimensional coordinate the radiation intensity variation at a fix distance into the direction (θ, φ). This diagram has to be independent to the distance, and depends only to the observation direction (θ, φ) [7]. We present in the Figure 6, the radiation diagram of our proposed UWB antenna at nine frequencies.

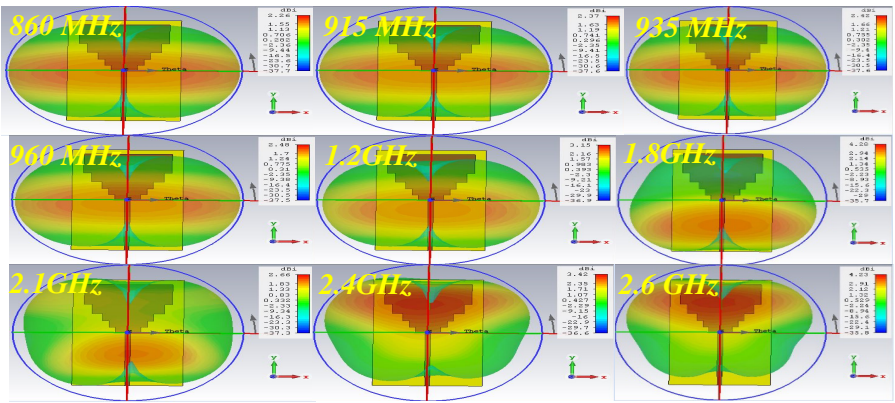

**Fig. 6.** Radiation diagram variation according with some landmark frequencies

## 2.5 Equivalent electronic circuit

The electrical equivalent circuit of this antenna must have the same response than the last simulation result by CST. For completing this task, we used the ADS software for representing the architecture of the electronic elements used. This circuit is formed in base of inductors, capacitors and resistors. The values of these electronic elements are given by theoretical equations, which are getting from **Gauss** theory (capacitor value equation 2), **Ampere** theory (inductor value equation 3) and the resonance condition (resistor value equation 4).

These equations are given according with the width (W), the length (d) and the thickness (h) of the resonator blocks of the antenna [8]. For our model, we have exactly six blocks, as shown in Figure 7.

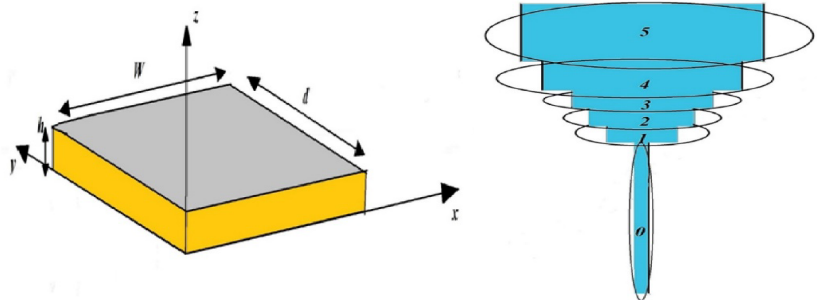

**Fig. 7**.     Antenna Resonator blocks

It's clearly seen, that the proposed antenna consists of six RLC resonators (Fig. 8). The resistor is due to the obstacle that the excitation current will encounter, and this obstacle is defined by the change in shape of the patch, the capacitor C is due to the spacing between the ribbon and the ground plane, the inductor L is due to the decrease in the width of the patch.

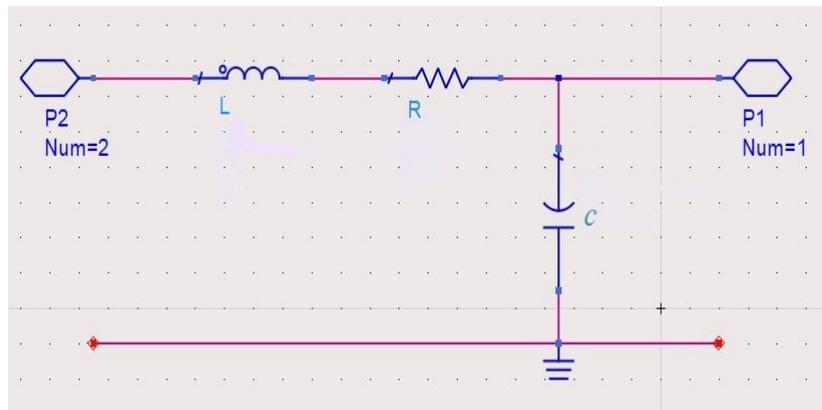

**Fig. 8.**     RLC resonator

The parameters C, L, R, are given by the following formulas [8].

$$C_{eq} = \frac{W\epsilon_0\epsilon_r}{d\cdot\left[\frac{\ln\left(\frac{2d(W+h)+2hW}{2hd}\right)}{d+h}+\frac{\ln\left(\frac{2d(W+h)+2hW}{2hW}\right)}{W+h}\right]} \cong n\cdot\frac{\epsilon_0\epsilon_r Wh}{50d^2} \quad (2)$$

$$L_{eq} = \frac{\mu_0 d}{10{,}5W}\left[d.\ln\left(\frac{W+d}{d}\right) + W.\ln\left(\frac{(W+d)^2}{hW}\right)\right] \quad (3)$$

$$R_{eq} = \frac{n}{10}\sqrt{\frac{L}{C}}|1 - LC\omega^2| \quad (4)$$

Where, n is the blocks number for assuring the continuity between all cavities, W and d and h are, the cavity width, the cavity length, the cavity thickness respectively, $\epsilon_r$ is the relative permittivity, $\epsilon_0$ is the vacuum permittivity equal to $8.85.10^{-12} F/m$, and finally $\mu_0 = 4.\pi.10^{-7}$ H/m is the vacuum permeability. So, for each antenna or cavity futures, we deduce an electrical equivalent circuit, where the elements values (Table. 2) are calculated by using the previous equations (2, 3 and 4).

**Table. 2** The electronic element values for each cavity

|  | **W(mm)** | **d(mm)** | **$C_{eq}$(pF)** | **$L_{eq}$(nH)** | **$R_{eq}$(Ω)** |
|---|---|---|---|---|---|
| **Cavity 0** | 3.2 | 60 | 0.0014 | - | - |
| **Cavity 1** | 18 | 7 | 0.417 | 2.39 | 3.3 |
| **Cavity 2** | 27 | 7.5 | 1.09 | 3.28 | 22.35 |
| **Cavity 3** | 36 | 8.5 | 1.69 | 3.9 | 20 |
| **Cavity 4** | 51 | 10.5 | 2.1 | 5.15 | 37.5 |
| **Cavity 5** | 62 | 24 | 4.2 | 10 | 35.5 |

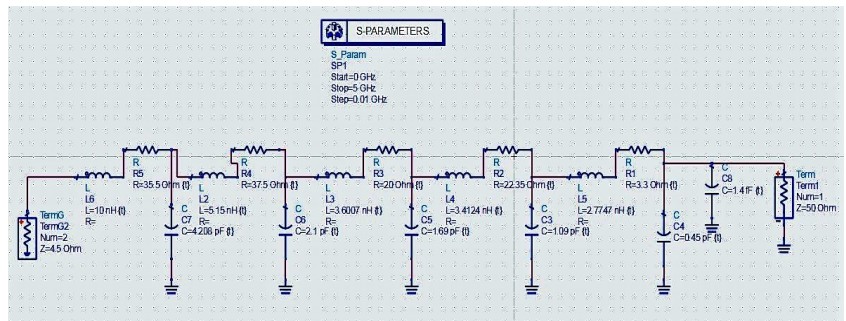

**Fig. 9.**   Electrical equivalent circuit of the antenna

Figure 9 shows the electrical equivalent circuit under ADS software, Figure 10 presents two graphs of $S_{11}$ of this antenna simulated by CST and that of the circuit by ADS, in order to make a comparison.

The first side, located in the right of the Figure 9, is the input impedance with 50Ω value, but in the other side the impedance is taken 4.5 Ω, this impedance value is calculated from the next equations (5, 6, 7, 8) according with the antenna parameters width (W), thickness (h), effective permittivity $\epsilon_{eff}$) [9]:

**When $\left(\frac{W}{h}\right) < 1$**

$$\epsilon_{eff} = \frac{\epsilon_r + 1}{2} + \frac{\epsilon_r - 1}{2}\left[\left(1 + 12 \cdot \left(\frac{h}{w}\right)\right)^{-\frac{1}{2}} + 0.04\left(1 - \left(\frac{w}{h}\right)\right)^2\right] \quad (5)$$

$$z_0 = \frac{60}{\sqrt{\epsilon_{eff}}} \ln\left(8\frac{h}{w} + 0.25\frac{w}{h}\right) \quad (6)$$

**When $\left(\frac{W}{h}\right) \geq 1$**

$$\epsilon_{eff} = \frac{\epsilon_r + 1}{2} + \frac{\epsilon_r - 1}{2}\left(1 + 12 \cdot \left(\frac{h}{w}\right)\right)^{-\frac{1}{2}} \quad (7)$$

$$z_0 = \frac{120.\pi}{\sqrt{\epsilon_{eff}} \cdot \left[\frac{w}{h} + 1.393 + \frac{2}{3}\ln\left(\frac{W}{h} + 1.444\right)\right]} \quad (8)$$

For the last cavity, the width is equal to 62 mm, the length is 24 mm, and 1.7 mm for the thickness, we have also $\left(\frac{W}{h}\right) \geq 1$, so the effective permittivity is equal to 4.2, and $z_0 = 4.5\Omega$.

### *2.6 Comparison and discussion*

The equivalent electronic circuit elements are given according with the antenna dimension parameters, so its response has to be the same as the simulation one. The Figure 10, shows that both responses are too much closers and we can say that they are duplicated within the frequency range [0.8 – 4.1] GHz with an $S_{11}$ value less than -15 dB.

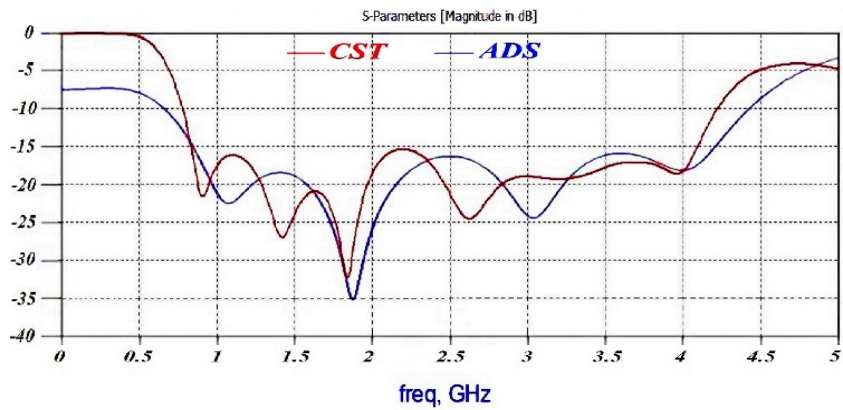

**Fig. 10.** Antenna (CST) and circuit simulation (ADS) responses

## 3  UWB antenna realization and practical results

### 3.1  The proposed model

The Figure 11 shows the realized antenna with a SMA excitation port such its impedance is fixed at 50 Ohm, the Figure 11 (a) presents the top side and the Figure 11 (b) the bottom side of this UWB antenna.

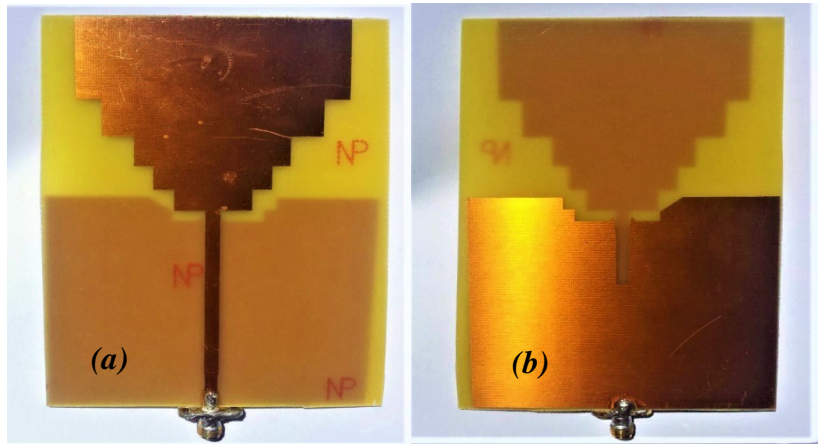

**Fig. 11.**  (a) Top side, (b) bottom side of the proposed antenna

### 3.2  Measured return loss

By using the network analyzer, we deduce the measured $S_{11}$ parameter of this antenna, depicted in the Figure 12.

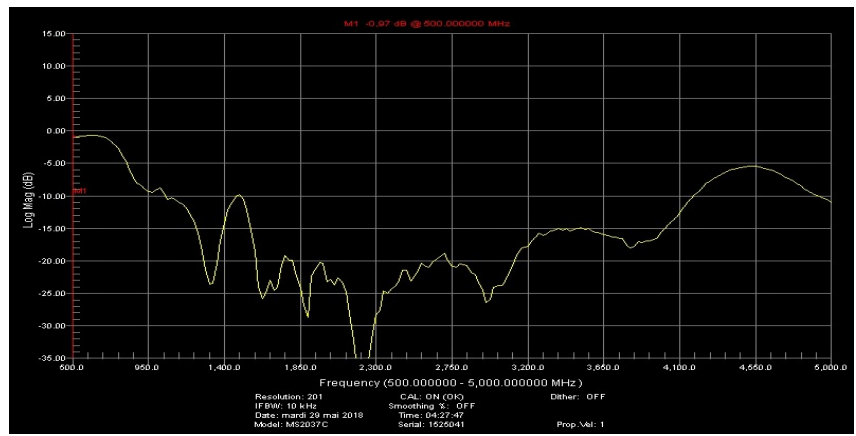

**Fig. 12.**  Practical $S_{11}$ response

### 3.3 Comparison and discussion

The Figure 13 allows to make a comparison between the measured and the simulated results and then to discuss the similarity of these both results.

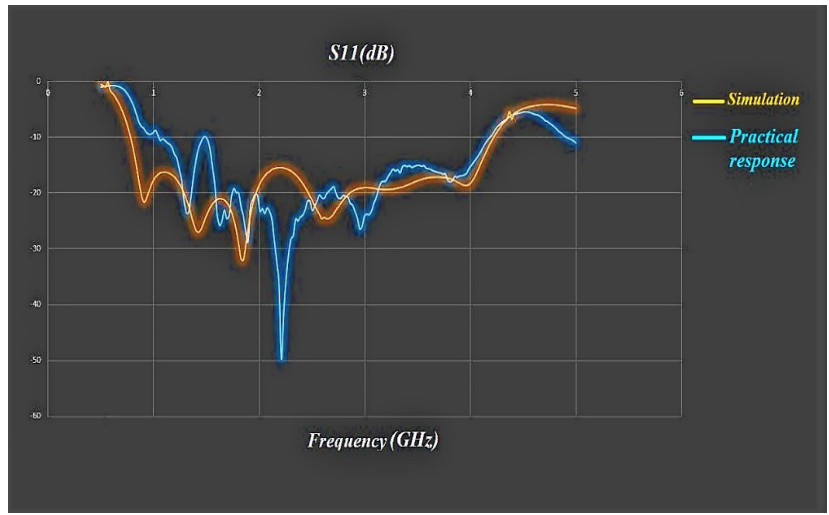

**Fig. 13**. Simulation and the practical response ($S_{11}$) of the antenna

As we can see, both responses are too similar, with a similarity rate of around 94%, the major difference is located in the frequency range (0.78 – 1.1) GHz, a second difference is around 2 GHz, due to the imperfection of the material used for the realization of this antenna.

## 4 Conclusion

We have designed, simulated and realized a UWB patch antenna on FR4 substrate hybrid technology. The performances of the antenna are studied using CST and ADS software's. The obtained results show that the simulations and the measurements are in good agreement. the simulated and the measured $S_{11}$ responses are in good concordance inside the frequency bandwidth [0.78 – 4.2] GHz.
With a measured bandwidth defined from 980MHz to 4.2GHz, this antenna could be used for a large domain of applications, such as radars, mobile communications, wireless network applications…

**References**

1. J. L. Volakis, Antenna Engineering Handbook, USA, NY, New York:McGraw-Hill, 2007.
2. D. Ramaccia, F. Bilotti, A. Toscano, and L. Vegni, "Dielectric-free multi-band frequency selective surface for antenna applications", COMPEL - The International Journal for Computation and Mathematics in Electrical and Electronic Engineering, Vol. 32, No. 6, pp. 1868-1875, 2013.
3. D. Ramaccia et. al. "Experimental Verification of Broadband Antennas loaded with Metamaterials" 2015 IEEE Antennas and Propagation Symposium, pp.1736-1737, 19 July 2015, Vancouver, CA.
4. T. Zasowski, F. Althaus, M. Stger, A. Wittneben, G. Trster, "UWB for noninvasive wireless body area networks: Channel measurements and results", IEEE Ultra-Wideband Syst. Technol. Conf. (UWBST 2003), 2003-Nov.
5. Bo Gao and Ge Wu and Jia-Yu Huo and Xiao-Jian Tian, "Planar Antenna Aids UWB Communications", microwaves & RF, November 2013.
6. Ch. Raghavendra, M. Suma and A. V. Akhila Krishna "Design and Analysis of Circular Patch Antenna for UWB Applications", Indian Journal of Science and Technology, Vol 9(S1), DOI:10.17485/ijst/2016/v9iS1/107887, December 2016.
7. Dominic Grenier,"Antenne et propagation radio", Université Laval Québec, Canada G1V 0A6, Hiver 2015, 439P.
8. Chemseddine Benkalfate, « Investigations sur les systèmes de collecte d'énergie RF et Micro-onde.», mémoire de fin d'étude, Université de Tlemcen, 2018, 125P.
9. Orfanidis, Sophocles J., "Electromagnetic Waves and Antennas Waveguides", Department of Electrical and Computer Engineering, Rutgers University, June 21, 2004, 1074P.